\documentclass[prd,twocolumn,superscriptaddress,floatfix]{revtex4-1}
\usepackage[utf8]{inputenc}
\usepackage{graphicx}
\usepackage{graphics,bm}
\usepackage{amssymb}
\usepackage{color}

\newcommand{\bra}[1]{\langle #1 |}
\newcommand{\ket}[1]{| #1 \rangle}

\def\6{\langle}
\def\9{\rangle}

\newcommand\al{\alpha}

\usepackage{amsmath}
\usepackage{amsfonts}
\usepackage{amssymb}

\newcommand\ta{\mathtt{a}}

\newcommand\tPsi{\mathtt{\Psi}}

\newcommand\wtPsi{\widetilde{\mathtt{\Psi}}}
\newcommand\wtPi{\widetilde{\mathtt{\Pi}}}
\newcommand\wtu{\widetilde{\mathtt{u}}}
\newcommand\wta{\widetilde{\mathtt{a}}}

\newcommand\tPA{\mathtt{\Psi}_{\scriptscriptstyle KG}^{\scriptscriptstyle A}}

\newcommand{\be}{\begin{equation}}
\newcommand{\ee}{\end{equation}}
\newcommand{\ba}{\begin{eqnarray}}
\newcommand{\ea}{\end{eqnarray}}

\usepackage{amsmath}

\begin{document}
\title{Unruh effect as a result of quantization of spacetime}
\author{Lucas C. C\'{e}leri}
\email{lucas@chibebe.org}
\affiliation{Instituto de F\'{i}sica, Universidade Federal de Goi\'{a}s, Caixa Postal 131, 74001-970, Goi\^{a}nia, Brazil}

\author{Vasileios I. Kiosses}
\email{kiosses.vas@gmail.com}
\affiliation{Instituto de F\'{i}sica, Universidade Federal de Goi\'{a}s, Caixa Postal 131, 74001-970, Goi\^{a}nia, Brazil}

\begin{abstract}
A way to encode acceleration directly into fields has recently being proposed, thus establishing a new kind of fields, the accelerated fields. The definition of accelerated fields points to the quantization of space and time, analogously to the way quantities like energy and momentum are quantized in usual quantum field theories. Unruh effect has been studied in connection with quantum field theory in curved spacetime and it is described by recruiting a uniformly accelerated observer. In this work, as a first attempt to demonstrate the utility of accelerated fields, we present an alternative way to derive Unruh effect. We show, by studying quantum field theory on quantum spacetime, that Unruh effect can be obtained without changing the reference frame. Thus, in the framework of accelerated fields, the observational confirmation of Unruh effect could be assigned to the existence of quantum properties of spacetime.
\end{abstract}

\maketitle

\noindent
\textbf{Introduction} --- While quantum mechanics arose from experimental observations which could not be reconciled with classical physics, till now, there is no experimental fact indicating that space and time are quantized. This is because the standard model, which describes all matter we have observed, is based on flat space quantum field theory and general relativity, which describes gravity, takes no account of quantum theory. 

In general relativity, the gravitational field and the spacetime are the same physical entity. All fields we know exhibit quantum properties at some scale, thus we believe that spacetime must have quantum properties as well.
The idea of quantizing spacetime dates back to the early days of quantum theory as a way to eliminate infinities from quantum field theories \cite{Snyder,Carazza} or just to survey the consequences of this assumption \cite{Hill}.

The first argument suggesting that due to gravitational field, spacetime breaks down at very short distances was advanced by Bronstein \cite{Bronstein}. Today, while it is generally accepted that spacetime is quantized, there is disagreement as to how quantization manifests itself \cite{Smolin}. Currently, the main candidates for a theory of quantum gravity are loop quantum gravity and string theory. Loop quantum gravity was the first major theory to postulate that spacetime must be quantized \cite{Rovelli}. String theory, in a different set, eventually came to the same conclusion \cite{Amati}.

Recently, a radical different approach was proposed by the authors, leading to the quantization of space and time by considering not gravity but acceleration \cite{C-K}. As is well known, Einstein initially used accelerated systems as a guide to understand the nature of gravity in classical physics. We believe this can also be done for quantum theory. In this vein, accelerated fields were introduced. The construction of accelerated fields differs from ordinary quantum field theory and provides a mathematically consistent way to quantize
spacetime. In this approach, space and time are quantized in the way quantities like energy and momentum are quantized in ordinary quantum field theories.

The theory of accelerated fields, as presented at \cite{C-K}, is an effort to quantize spacetime, while it is kept separated from the matter fields. Strictly speaking, the aim was only to locally describe the quantum behavior of the gravitational field. In this article, we proceed a step further and investigate the effects of accelerated fields on the Klein-Gordon massive field, by defining accelerated Klein-Gordon massive fields and comparing them with the inertial ones. As we will see below, the definition of accelerated Klein-Gordon fields actually gives rise to creation and annihilation operators that encapsulate noninertial effects. Spacetime becomes dynamical and affects the matter fields defined on it. The new creation and annihilation operators defined by this theory imply a modification of the usual commutation relation. Specifically, the new commutation relation is divided in two parts, the first one corresponding to the standard commutation relation while the second one reflects the quantum nature of spacetime. We thus find that space and time, due to acceleration, ought to be described as a non-commutative quantity. There has been much work recently studying consequences of formulating quantum mechanics on non-commutative manifolds in an attempt to describe quantum gravity \cite{S-W, Connes}. In this work we show that acceleration can be employed to develop theories of deformed commutation algebras as well.   

Moreover, considering the vacuum state of the accelerated Klein-Gordon field we find that it is equivalent to an inertial Klein-Gordon field in thermal equilibrium at temperature $T$. In other words, we find that in the presence of the accelerated field, an inertial observer should detect a flux of thermal particles whose temperature is proportional to the acceleration defined by the accelerated field.

The primary significance of Unruh effect \cite{Unruh76} lies in the argument that the very concept of particles is frame-dependent, demonstrating that the key features of Hawking radiation exists without gravity. Here we show that it is possible to derive Unruh effect without changing the reference frame, by just studying quantum field theory in the presence of the accelerated field. In this way, Unruh effect acquires its own source, responsible for the observable particle creation rate, in the same spirit as Hawking effect has the gravitational field in the form of Black holes \cite{Hawking74,Hawking75}. We observe that Unruh and Wald \cite{U-W} made clear that Unruh effect is compulsory for the consistency of field theory in accelerated frames. The main result of this paper is that Unruh effect is necessary in order to describe the consequences of formulation field theories on quantum space.   

It is interesting to note that in the context of string theory there are some works on quantum space which address Unruh effect \cite{Freidel-1,Freidel-2,Freidel-3}. In these works, the authors were interested in defining a quantum system with a built-in length scale. A quantum space were defined as a choice of polarization for the Heisenberg algebra of quantum theory. The derived quantum space employed in the present work, on the other hand, is based on Schrödinger quantum mechanics and relies on the classical notion of momentum appearing as arguments of the wave function \cite{C-K}.

In \cite{C-K} it has been shown a way to build particle states characterized by uniform acceleration. In order to set notation and the necessary mathematical background, in the following we briefly summarize the main elements of this structure. After this, we proceed on its generalization by including states which describes accelerated Klein-Gordon particles, and state our main result. We consider quantum field theories in two dimensions with metric signature $+-$. Furthermore, we shall use units such that $\hbar=c=k=1$, unless specified otherwise. 

\vspace{0.2cm}
\noindent
\textbf{background} --- Let us start by considering a real scalar accelerated field $\wtPsi(p^\mu)$ in momentum space, defined by the wave equation \cite{C-K}
\be
\left({\widetilde{\partial}_E}^2-{\widetilde{\partial}_p}^2 - \frac{1}{\alpha^2} \right)\wtPsi(p^\mu) = 0,
\label{eq.wem}
\ee
with $p^{\mu} = (E,p)$ being the four momentum while $\alpha$ represents the proper acceleration. From here on, quantities denoted with (without) a tilde are associated with Eq. (\ref{eq.wem}) (the Klein-Gordon equation). There is a clear similarity between Eq. (\ref{eq.wem}) and the Klein-Gordon equation. However there is also an important difference: For accelerated fields the wave equation is space-like. The way to deal with this discrepancy is to solve the theory by introducing an orthonormal set of mode solutions given by $\wtu_t(p^\mu) = e^{i(E t - p x_t)}/\sqrt{4 \pi x_t}$, with $x_t \equiv x(t) = \sqrt{t^2 + 1/\al^2}$, and claiming that the definition of the positive- and negative-frequency solutions lies in the existence of a space-like Killing vector field, $\widetilde{\partial}_p$, in momentum space \cite{C-K}.

Working in the space spanned by the positive frequency modes $\wtu_t$, one can define the field operator $\wtPsi$ in the language of the second quantization in the usual way
\be
\wtPsi(p^\mu) = \int dt \left( \wta_t \wtu_t(p^\mu) + \wta_t^\dagger \wtu_t^*(p^\mu)\right).
\label{eq.FOM}
\ee
$\wta_t$ ($\wta_t^\dagger$) is the annihilation (creation) operator, which acts on coordinate space and annihilates (creates) excitations at space-time point ($t,x_{t}$). From these operators, the field Hamiltonian takes the form
\be
\widetilde{H} = \int \frac{dt}{\sqrt{4\pi x_t}} \,  x_t\, \wta_{t}^\dagger\,\, \wta_{t}.\label{eq.HA}
\ee
The Hamiltonian was defined in the following way. First we construct the Lagrange density associated with equation (\ref{eq.wem}) by inverting the Euler-Lagrange equation. Then, Noether's theorem provides us the momentum independent quantity that plays the role of the Hamiltonian. Finally, by using Eq. (\ref{eq.FOM}) we can rewrite the Hamiltonian in the form shown in Eq. (\ref{eq.HA}).

By defining the conjugate momentum as $\wtPi_E(p) = \widetilde{\partial}_p \wtPsi_E(p)$, and postulating the canonical \emph{equal-momentum} commutation relations
\ba
\left[\wtPsi_p(E),\wtPsi_p(E')\right]&=& \left[\wtPi_p(E),\wtPi_p(E')\right] =0, \nonumber 
\\
\left[\wtPsi_p(E),\wtPi_p(E')\right]&=& i \, \delta(E-E'),\nonumber 
\ea
we get
\be
\left[\widetilde{\mathtt{a}}_{t},\widetilde{\mathtt{a}}_{t'}\right] = \left[\widetilde{\mathtt{a}}_{t}^\dagger,\widetilde{\mathtt{a}}_{t'}^\dagger\right] = 0\quad \mbox{and} \quad \left[\widetilde{\mathtt{a}}_{t},\widetilde{\mathtt{a}}_{t'}^\dagger\right] = \delta(t-t').
\label{crmCAO}
\ee

Since the individual solutions $\wtu_t$ are associated with positive frequency, and the Hamiltonian is a sum over the contributions from each $t$ value, there will be a single vacuum state $\ket{0_{\wta}}$, defined by $\wta_t \ket{0_{\wta}} =0$. The excitations will then be defined by $\wta_t^\dagger \ket{0_{\wta}}$ and are interpreted as single accelerated particle. In the usual quantum field theory every excitation is said to carry a specific energy and momentum. In our case, since the wave equation is defined in the momentum space, we say that every excitation will \emph{carry} a specific $t$ and $x_{t}$.

\vspace{0.2cm}
\noindent
\textbf{Results} --- Having defined the accelerated fields in momentum space, we return now to coordinate space in order to discuss the ordinary theory of quantum fields in the presence of an accelerated quantum field. We consider the simplest possible case, a real massive scalar field $\Phi$ in Minkowski space-time obeying the massive Klein-Gordon equation $(\partial_t^2 - \partial_x^2 + m^2)\Phi(t,x) = 0$. The normal-mode solutions to this equation, in Minkowski coordinates, are $\phi_p(x^\mu) = e^{-i(E_p t - p x)}/\sqrt{4 \pi E_p}$, with $E_p= \sqrt{p^2 + m^2}$. The Klein-Gordon inner product, which is expressed as an integral over a constant-time hypersurface, allow us to define the annihilation operator associated to $\phi_p$ by $a_p = \left( \phi_p, \Phi \right)_{KG}$. From this we can write
\be
\Phi(t,x) = \int dp \left(a_p \phi_p(t,x) + a^\dagger_p \phi_p^*(t,x) \right),\nonumber 
\ee
whose associated vacuum state $\ket{0_{I}}$ is defined by $a_p \,\ket{0_{I}} =0$. The states $a_p^\dagger \ket{0_{I}}$ are interpreted as single particle states with momentum $p$ and energy $E_p$,  

Computing the action of $\Phi(t,x)$ on $\ket{0_{I}}$ one finds 
\be
\Phi(t,x)\ket{0_{I}} = \int dp \,  \phi_p^*(t,x) \, a^\dagger_p \ket{0_{I}}\nonumber 
\ee
which corresponds to the superposition of single particle momentum eigenstates and thus, corresponds to a particle at $(t,x)$.

The theory of accelerated fields, as we have seen, has been established in momentum space, thus the basic degrees of freedom are operator valued functions of energy and momentum, i.e. the field operator $\wtPsi(p^\mu)$ and its conjugate momentum $\wtPi(p^\mu)$. However, writing the wave equation (\ref{eq.wem}) in Fourier space (which actually is the coordinate space), each Fourier mode of the field is treated as an independent oscillator with its own annihilation and creation operator. Hence, from Eq. (\ref{eq.FOM}) the Fourier transform of the field $\wtPsi (p^\mu)$ reads
\be
\wtPsi(t) = \wta_{t} \wtu_{t}(p^\mu) + \wta_{t}^\dagger \wtu^*_{t}(p^\mu). \nonumber 
\ee
The operator $\wta_t^\dagger$ ($\wta_{t}$) creates (annihilates) particles associated with the mode functions $\wtu_{t}(p^\mu)$. For a given time $t$, there are two plane wave solutions for $\wtu_{t}$, one with positive $x_t$ and one with negative $x_t$. This is similar to the solutions for the Klein-Gordon equation for a specific momentum, composed of positive and negative energies. Therefore, the set of solutions of the Eq. (\ref{eq.wem}) are distributed into two space-like separated regions.

Considering the Klein-Gordon field operator $\Phi(t,x)$, we impose the equal-time canonical commutation relation $\left[\Phi(t,x),\Pi(t,x')\right] =i \delta(x-x')$, with $\Pi(t,x')$ being the momentum conjugate to $\Phi(t,x)$. Moreover, we add the relation
\[
\left[\Phi(t,x),\Phi^\dagger (t',x) \right] = \delta(t-t').   
\]

The algebra stemmed from the above commutation relation, typical for creation and annihilation operators, are identical to those we found for the quantized accelerated field by using plane waves as an expansion basis. Thus, writing 
\be
\xi_{\scriptscriptstyle KG}^{\scriptscriptstyle A}(p^\mu) =\int dt \left( \Phi_{t} \wtu_{t}(p^\mu) + \Phi_{t}^\dagger \wtu^*_{t}(p^\mu) \right), \nonumber 
\ee
the operators $\Phi_{t}^\dagger$ and $\Phi_{t}$ acquire a dual role. As fourier coefficients in the expansion of the wave operator $\xi_{\scriptscriptstyle KG}^{\scriptscriptstyle A}$ they create and annihilate particles associated with the mode function $\wtu_{t}$ and, as a Klein-Gordon field operator, they create Klein-Gordon particles at $(t,x_t)$. Note that $\xi_{\scriptscriptstyle KG}^{\scriptscriptstyle A}(p^\mu)$ is a combination of quantities with and without the tilde. 

Now, the field 
\be
\tPsi_{\scriptscriptstyle KG}^{\scriptscriptstyle A}(t) = \Phi_{t} \wtu_{t}(p^\mu) + \Phi_{t}^\dagger \wtu^*_{t}(p^\mu).\label{eq.AKGF}
\ee
obeys the Klein-Gordon equation. Since we are in Minkowski spacetime, the time-translation symmetry allows us to unambiguously define a set of positive- and negative-frequency modes, whose temporal behavior is given by $e^{-i E_p t}$ and $e^{i E_p t}$, respectively, and to expand the field operator $\tPsi$ in terms of these modes. Choosing the pair $\{\phi_p,\phi_p^*\}$, which are expressed in Minkowski coordinates, the expansion of $\tPsi$ in terms of these modes reads
\be
\tPsi_{\scriptscriptstyle KG}^{\scriptscriptstyle A}(t) = \int dp \left(\ta_p^A \phi_p(t,x_t) + \left(\ta^A_p\right)^\dagger \phi_p^*(t,x_t) \right). \label{eq.30}
\ee
with the operator coefficients, $\left(\ta^A_p\right)^\dagger$ and $\ta_p^A$, to be interpreted, as usual, as creation and annihilation operators with respect to the set of modes $\{\phi_p,\phi_p^*\}_{x \rightarrow x_t}$. In the standard way, $\tPsi_{\scriptscriptstyle KG}^{\scriptscriptstyle A}(t)$ is represented as an operator-valued distribution in a Fock space spanned by the vacuum state $\ket{0_A}$, defined by $\ta_p^A \ket{0_{A}} =0$ and by the (unormalized) $n$-particle states of the form $\left( \ta^A_{p_1}\right)^\dagger\, \left(\ta^A_{p_2}\right)^\dagger \ldots \left(\ta^A_{p_j}\right)^\dagger\, \ket{0_{A}}$.

The excitations defined by $\left(\ta^A_{p}\right)^\dagger \ket{0_{A}}$ carries energy $\sqrt{p^2 + m^2}$ while the ones associated with $\tPsi_{\scriptscriptstyle KG}^{\scriptscriptstyle A}(t) \ket{0_{A}}$ are located at position $\sqrt{t^2 + 1/\al^2}$. Thus, the field operator is associated with single particles with mass $m$ and acceleration $\al$. In this space the (renormalized) Hamiltonian is given by
\be
H^A = \int d p \, E_p \left( \ta^A_{p}\right)^\dagger \, \ta_p^A \nonumber 
\ee
where the operator $N_A = \left( \ta^A_{p}\right)^\dagger  \ta_p^A$ counts the number of accelerated particles with momentum $p$.

The zero-point fluctuations that a quantum harmonic oscillator exhibits arise formally from the non-commutativity of the corresponding creation and annihilation operators. Like a harmonic oscillator, if we consider the field component 
\be
\tPsi_{\scriptscriptstyle KG}^{\scriptscriptstyle A}(t) = \Phi_{t} \wtu_{t}(p^\mu) + \Phi_{-t}^\dagger \wtu^*_{-t}(p^\mu),\nonumber
\ee
which has zero mean in the vacuum state $\ket{0_I}$, it must undergoes zero point fluctuations due to the non-commutativity of $\Phi_{t}$ and $\Phi_{t'}^\dagger$,
\be
\bra{0_I} \left(\tPsi_{\scriptscriptstyle KG}^{\scriptscriptstyle A}(t)\right)^\dagger\, \tPsi_{\scriptscriptstyle KG}^{\scriptscriptstyle A}(t) \ket{0_I} = |\wtu_{-t}(p^\mu)|^2 = \frac{1}{4 \pi x_t}. \label{eq.39}
\ee

The meaning of this equation is that, at every point in momentum space, characterized by a specific acceleration ($x_t$ is acceleration dependent), a pair of spatially separated accelerated particles comes out in coordinate space. This is similar to the usual quantum field theory, when vacuum fluctuations induce the creation of a time-like pair. 

We found that the relationship between $\tPsi_{\scriptscriptstyle KG}^{\scriptscriptstyle A}(t)$ and $\Phi(t,x)$ is given by (\ref{eq.AKGF}). Since both fields were quantized using the normal Minkowski modes, their difference reduces to the difference between their corresponding creation and annihilation operators. In the case of real-valued Klein-Gordon fields, i.e. Hermitean field operators $\Phi^\dagger = \Phi$, the Klein-Gordon inner product of the field operators $\tPsi_{\scriptscriptstyle KG}^{\scriptscriptstyle A}(t)$ and $\phi_{p}(t,x)$ ($\phi_{p}^*(t,x)$) gives the Fourier coefficient $\ta_p^A$ ($\left(\ta^A_p\right)^\dagger$), and, due to Eq. (\ref{eq.AKGF}), we have 
\be
\ta_p^A = a_p \left(\wtu_{t}^* +\frac{\wtu_{t}+\wtu_{t}^*}{2}\right) + a^\dagger_{-p} \left(\wtu_{t}^*-\frac{\wtu_{t}+\wtu_{t}^*}{2}\right)e^{i 2 E_p t}\label{eq.85}
\ee
where the wave functions $\wtu_{t}$ are taken at $E=E_p$.
The commutation relations for $\ta_p^A$ and $(\ta_p^A)^\dagger$ can be found to be 
\be
[\ta_p^A, (\ta_{p'}^A)^\dagger] = 2 [a_p,a^\dagger_{p'}] \left(1 + \frac{\wtu^2_{t}+ (\wtu^*_{t})^2}{2}\right).
\ee 
As we see, the presence of the accelerated wave modes in the solution of the Klein-Gordon equation modifies the commutation relation between creation and annihilation operators.

The transformation (\ref{eq.85}), since it mixes annihilation and creation operators, is actually the Bogolubov transformation. Thus, the vacuum states defined by the inertial annihilation operator $a_p$ and the accelerated annihilation operator $\ta_p^A$ are distinct due to the presence of $\wtu_t$ in the latter. More specific, the expectation value of the  number operator $N_A$ for the ground state of the inertial free theory $\ket{0_I}$ is given by
\be
\bra{0_I} N_A \ket{0_I} =\frac{1}{2}\left(1-\wtu^2_{t}- (\wtu^*_{t})^2\right)= \frac{\sin^2[(E_p t - p x_t)]}{4 \pi x_t}, \nonumber 
\ee 
which does not vanish for every $p$.

Until now we have shown the effect of accelerated fields on the commutation relations of an inertial field theory and the disagreement between an inertial- and an accelerated Klein-Gordon fields on the definition of the ground state. Now we will show that this difference is due to the thermal behavior of the accelerated field. In order to show this, we derive a relation between the inertial description of an inertial- and an accelerated Klein-Gordon field. The approach presented below is based on Ref. \cite{Miloni} where the thermal effect of acceleration was computed by means of the field correlation function.

We work with the inertial Klein Gordon field $\Phi(t,x)$ and the accelerated one $\tPA(t)$, as defined in the previous section, with the only difference being that here it is assumed to be massless, in order to simplify the calculations.

Starting with $\Phi(t,x)$, consider the two-time correlation function $\left\langle \Phi(t,0)\, \Phi(t+\tau,0)\right\rangle$ at a specific point in space for the field in thermal equilibrium at temperature $T$, i. e. the mode whose frequency is $E_p$ has $n(E_p)$ particles on average
\be
\left\langle a_{p'}^\dagger \, a_{p}\right\rangle = \delta(p'-p)\, n(E_p),\qquad n(E_p)=\left(e^{E_p/T}-1\right)^{-1}. \nonumber
\ee
From this it follows that
\be
\left\langle \Phi(t,0)\, \Phi(t+\tau,0)\right\rangle = -\frac{1}{\pi} (\pi T)^2 \texttt{csch}^2 \left(\pi T \tau\right) \label{eq.45}
\ee

We proceed with $\tPA(t)$, considering the correlation function $\left\langle \tPA(t_1)\,\,\tPA(t_2)\right\rangle_{0_A}$ in the vacuum state $\ket{0_A}$. In this case it holds $\left\langle \ta_p^A \, \ta_{p'}^A\right\rangle_{0_A} = \left\langle (\ta_p^A)^\dagger (\ta_{p'}^A)^\dagger\right\rangle_{0_A} = \left\langle (\ta_{p}^A)^\dagger \,\ta_{p'}^A\right\rangle_{0_A} =0$ and $\left\langle \ta_p^A (\ta_{p'}^A)^\dagger\right\rangle_{0_A} = \delta(p-p')$. Therefore, from Eq. (\ref{eq.30}), we obtain
\be
\left\langle \tPA(t_1)\,\,\tPA(t_2)\right\rangle_{0_A} = \frac{1}{\pi}\frac{1}{\Delta x^2 - \Delta t^2}. \nonumber
\ee
with $\Delta t = t_2 -t_1$ and $\Delta x = x_{t_2} - x_{t_1}$. By construction, the relation $x_t^2 - t^2 = 1/\al^2$ holds. As this represents hyperbolic curves, we can use hyperbolic functions and set
\ba
x_{t_i} &=& \frac{1}{\al} \cosh(\al \rho_i)            \label{eq.47} \\
t_i &=& \frac{1}{\al} \sinh(\al \rho_i), \qquad i=1,2    \label{eq.48}
\ea
with $\rho_i$ a parameter. One can calculate the difference ${\Delta x^2 - \Delta t^2}$ by making use of Eqs. (\ref{eq.47}) and (\ref{eq.48})
\be
\Delta x^2 - \Delta t^2 = -\frac{4}{\al^2} \sinh^2\left(\frac{\al(\rho_2-\rho_1)}{2}\right). \nonumber
\ee 
The correlation function $\left\langle \tPA(t_1)\,\,\tPA(t_2)\right\rangle_{0_A}$ in the vacuum of the accelerated Klein-Gordon field, as measured by an inertial observer, is given by 
\be
\left\langle \tPA(t_1)\,\,\tPA(t_2)\right\rangle_{0_A} = -\frac{\al^2}{4 \pi} \texttt{csch}^2\left(\frac{\al(\rho_2-\rho_1)}{2}\right)\nonumber
\ee
which is equivalent to the thermal-field correlation function (\ref{eq.45}) with temperature
\be
T = \frac{\al}{2\pi }. \nonumber
\ee 
The meaning of this result is that an inertial observer, being in the ground state of a quantum field, responds differently in the presence of an accelerated field. The effect of the accelerated field is to promote the zero-point quantum field fluctuations to the level of thermal fluctuations.

\vspace{0.2cm}
\noindent
\textbf{Conclusions} --- Particle creation in the case of Hawking effect for black holes and Unruh effect for accelerated observers are consequences of the application of the general framework of quantum field theories in curved spacetimes. Since, in a curved spacetime setting, there is no analog of the preferred Minkowski vacuum, the very notion of a particles becomes ambiguous. Our formulation avoid such difficulties because it does not rely on these ideas since we work in flat spacetime and considering only inertial reference frames. In our case, particles will be emitted from the vacuum of an inertial quantum field due to the presence of the accelerated field, that acts as a source for such effect.

Keeping in mind that the mathematical language we use here is that of quantum field theory and not that of string theory, our formulation integrates some of the ideas developed in metastring theory \cite{Freidel-4}. The habitat of metastrings is a form of quantum spacetime which reveals a geometric structure behind quantum theory. In that structure spacetime and momentum space appear to hold equal parts. Sharing the same view, we have promoted momentum space by constructing a relativistic field theory in momentum space analogously to standard field theories in spacetime.
	
The concept of Born reciprocity \cite{Born} simply states that physics should be equivalently formulated from the position and momentum point of views. In metastring theory, Born reciprocity is related to a fundamental symmetry of string theory, that of T-duality, which is generalized to Born duality \cite{Freidel-4}. In our formulation T-dualty reduces to a Fourier transform, which exchanges field operators in momentum space with field operators in spacetime. Precisely this existence of two types of quantum field theories in spacetime (the standard field theory and the Fourier transform of the accelerated field theory), interacting in the way presented in the present work, results in Unruh effect. While the first one, the standard field theory, is associated to the dynamics, the second one should be related to kinematics. In that case the vacuum is identified with a state that is annihilated by both the "dynamical" and "kinematical" annihilation operators. This is reminiscent of the Unruh effect where different vacua correspond to different reference frames. The analogy is established by identifying the accelerated observer's reference frame with the inertial one in the presence of accelerated fields i.e. the "kinematical" fields.

Even though gravity, as described in general relativity, appears not to be compatible with quantum field theory, this is not the case for the accelerated fields. The fact that the accelerated fields are quantized under the canonical procedure allowed us to precisely calculate the effect of acceleration on the inertial wave operator. In addition, we went one step further and argued that this change of the wave operator, as measured by an inertial observer, is responsible for the Unruh effect.

In this work we adopt the view that certain classical notions about space and time are inadequate at the fundamental level. We define accelerated Klein-Gordon fields by forming a notion of Klein-Gordon field being located with respect to the accelerated field. The presence of accelerated field modifies certain classical notions about space and time \cite{C-K}. Thus, this approach provides a profound conceptual shift in the understanding of Unruh effect. The vacuum of an accelerated quantum field theory is filled with particles, not because the concept of the number of particles is frame dependent, but because the classical notions of space and time are changed. 

Although Unruh thermal bath is accepted by the majority of physicists, the phenomenon still lacks observational confirmation \cite{Matsas}. Recently, an experiment based on classical physics was proposed to confirm the existence of Unruh effect \cite{Matsas2}. The analysis presented in this work represents a new path to obtain physically testable predictions of quantum spacetime. 

\vspace{0.2cm}
\noindent
\textbf{Acknowledgments} --- We would like to thanks George E. A. Matsas, Daniel Terno and Daniel A. T. Vanzella for discussions. We acknowledge financial support from the Brazilian funding agencies CNPq (Grants No. 401230/2014-7, 305086/2013-8 and 445516/2014-) and CAPES (Grant No. 6531/2014-08), the Brazilian National Institute of Science and Technology of Quantum Information (INCT/IQ).

\end{document}